\documentclass[fleqn,usenatbib]{mnras}

\usepackage{newtxtext,newtxmath}
\usepackage[T1]{fontenc}

\DeclareRobustCommand{\VAN}[3]{#2}
\let\VANthebibliography\thebibliography
\def\thebibliography{\DeclareRobustCommand{\VAN}[3]{##3}\VANthebibliography}

\usepackage{graphicx}	% Including figure files
\usepackage{amsmath}	% Advanced maths commands

\newcommand{\mmsun}{M/M_{\odot}}
\newcommand{\nuLnu}{\nu L_{\nu}}
\newcommand{\mdot}{\dot{m}}
\newcommand{\Mbh}{M_{\rm{BH}}}
\newcommand{\msun}{M_{\odot}}
\newcommand{\MMdotedd}{\dot{M}/\dot{M}_{\rm{Edd}}}
\newcommand{\LLedd}{L/L_{\rm{Edd}}}

\newcommand{\rev}{}
\newcommand{\revv}{} %second revision

\title[Collapse of the Disc]{Systematic Collapse of the Accretion Disc Across the Supermassive Black Hole Population}

\author[S. Hagen et al.]{Scott Hagen,$^{1}$\thanks{E-mail: scott.hagen@durham.ac.uk}
Chris Done,$^{1}$
John D. Silverman,$^{2,3, 4, 5}$
Junyao Li,$^{6}$
Teng Liu,$^{7}$
Wenke Ren,$^{8, 2}$
\newauthor
Johannes Buchner,$^{9}$
Andrea Merloni,$^{9}$
Tohru Nagao$^{10, 11}$
and Mara Salvato$^{9}$
\\
$^{1}$Centre for Extragalactic Astronomy, Department of Physics, Durham University, South Road, Durham, DH1 3LE, UK\\
$^{2}$Kavli Institute for the Physics and Mathematics of the Universe (Kavli IPMU, WPI), UTIAS, Tokyo Institutes for Advanced Study, University of Tokyo, \\ Chiba, 277-8583, Japan\\
$^{3}$Department of Astronomy, School of Science, The University of Tokyo, 7-3-1 Hongo, Bunkyo, Tokyo 113-0033, Japan \\
$^{4}$Center for Data-Driven Discovery, Kavli IPMU (WPI), UTIAS, The University of Tokyo, Kashiwa, Chiba, 277-8583, Japan \\
$^{5}$Center for Astrophysical Science, Department of Physics \& Astronomy, John Hopkins University, Baltimore, MD 21218, USA \\
$^{6}$Department of Astronomy, Univerity of Illinois at Urbana-Champaign, Urbana, IL 61801, USA \\
$^{7}$Department of Astronomy, University of Science and Technology of China, Hefei 230026, China \\
$^{8}$CAS Key Laboratory for Research in Galaxies and Cosmology, Department of Astronomy, University of Science and Technology of China, Hefei 230026, China \\
$^{9}$Max-Planck-Institute f\"{u}r Extraterrestrische Physik, Giessenbachstrasse 1, 85748, Garching bei M\"{u}nchen, Germany \\
$^{10}$Research Center for Space and Cosmic Evolution, Ehime University, 2-5 Bunkyo-cho, Matsuyama, Ehime, 790-8577, Japan \\
$^{11}$Amanogawa Galaxy Astronomy Research Center, Kagoshima University, 1-21-35 Korimoto, Kagoshima 890-0065, Japan
}

\date{Accepted XXX. Received YYY; in original form ZZZ}

\pubyear{\the\year{}}

\begin{document}
\label{firstpage}
\pagerange{\pageref{firstpage}--\pageref{lastpage}}
\maketitle

\begin{abstract}
The structure of the accretion flow onto supermassive black holes (SMBH) is not well understood. Standard disc models match to zeroth order in predicting substantial energy dissipation within optically-thick material producing a characteristic strong blue/UV continuum. However they fail at reproducing more detailed comparisons to the observed spectral shapes along with their observed variability. Based on stellar mass black holes within our galaxy, accretion discs should undergo a transition into an X-ray hot, radiatively inefficient flow, below a (mass scaled) luminosity of $\sim 0.02\,L_{\rm{Edd}}$. While this has been seen in limited samples of nearby low-luminosity active galactic nuclei (AGN) and few rare changing-look AGN, it is not at all clear whether this transition is present in the wider AGN population across cosmic time. A key issue is the difficulty in disentangling a change in spectral state from increased dust obscuration and/or host galaxy contamination, effectively drowning out the AGN emission. Here we use the new eROSITA eFEDS Survey to identify unobscured AGN from their X-ray emission, matched to excellent optical imaging from Subaru's Hyper Suprime-Cam; allowing the subtraction of the host galaxy contamination. The resulting, uncontaminated, AGN spectra reveal a smooth transition from a strongly disc dominated state in bright AGN, to the collapse of the disc into an inefficient X-ray plasma in the low luminosity AGN, with the transition occurring at $\sim 0.02\,L_{\rm{Edd}}$; revealing fundamental aspects of accretion physics in AGN.
\end{abstract}

\begin{keywords}
accretion, accretion discs -- black hole physics -- galaxies: active
\end{keywords}

%%%%%%%%%%%%%%%%%%%%%%%%%%%%%%%%%%%%%%%%%%%%%%%%%%

%%%%%%%%%%%%%%%%% BODY OF PAPER %%%%%%%%%%%%%%%%%%

\section{Introduction}

Active galactic nuclei (AGN) are powered by accretion onto a supermassive black hole (SMBH), but the structure of this flow is not well understood. Simple estimates of the accretion structure are that it should form a geometrically thin, relatively cool, disc if it is dense enough to thermalise \citep{Shakura73}, giving a disc temperature which increases with decreasing radius from the black hole. This predicts a spectral energy distribution (SED) consisting of a sum of black-body components, typically peaking in the UV/EUV for bright AGN. This matches to zeroth order to the strong blue/UV continuum seen in these sources, but more detailed comparisons fail \citep{Antonucci89, Lawrence18}. 

One issue is that AGN always show X-ray emission, with a non-thermal tail extending to high energies \citep{Elvis94, Lusso16}, indicating that some of the accretion energy must be dissipated outside of the optically thick disc structure. The X-ray emission is highly variable on short time-scales, indicating it originates in a compact coronal structure (e.g \citealt{Lawrence87, Ponti12b}) Compton scattering seed photons from the disc to higher energies. Additionally, even the bright blue/UV continuum does not match in detail to the disc models. There is a ubiquitous downturn in the UV which occurs before the expected disc peak \citep{Laor14, Cai23}, which appears to connect to an upturn in the X-ray spectra below $\sim 1$\,keV (generally referred to as the soft X-ray excess). These two features can be joined together across the unobservable EUV region to form an additional warm Comptonising component \citep{Laor97, Porquet04, Gierlinski04}, which cannot be explained by a either the standard disc model nor the hot Comptonising corona. 

Whatever the nature of the EUV component, this should carry most of the 
accretion power. EUV radiation is strongly ionising, producing characteristic broad emission lines where it irradiates nearby material (e.g \citealt{Wills85, Rees89, Peterson99, Kaspi05}). Current optical surveys take advantage of this when identifying and classifying AGN, by first selecting on the presence of a blue continuum in the photometry, and then confirming with spectroscopic follow up of broad emission lines (e.g \citealt{Richards02}). These optically selected AGN span a wide range in mass, and so any given luminosity bin will span a wide range in Eddington ratio. However, controlling for black hole mass shows a sharp drop in the number of objects with $\mdot \lesssim 0.01$ (e.g \citealt{Trump11, Mitchell23}), where $\mdot = \MMdotedd = L/L_{\rm{Edd}}$ (i.e. with the accretion efficiency factor included). This has generally been attributed to selection effects such as an increase in dust obscuration or host galaxy contamination, effectively hiding the AGN emission (e.g \citealt{Fabian08, Vasudevan13, Hickox18}). 

However, an alternative scenario is a change in the accretion structure, where the disc evaporates into a hot X-ray plasma, as often seen in galactic black hole binaries (BHBs) \citep{Done07}. These objects show strong thermal disc emission, now with temperatures peaking in the X-ray ($\sim 10^{7}$\,K), but only for
$\mdot \gtrsim 0.02$ \citep{Maccarone03}. 
Below this there is a dramatic spectral transition, from being dominated by soft (low energy) X-ray thermal emission, to being dominated by harder (high energy) Comptonised emission from significantly hotter $\sim 10^{9}$\,K X-ray plasma \citep{Maccarone03, Done07}. These hard spectra are best described by an alternative solution to the steady state accretion flow equations, one that is hot, optically thin and geometrically thick rather than the cool, optically thin, geometrically thin accretion discs. The best known of these alternative solutions is the Advection Dominated Accretion Flow (ADAF: \citealt{Narayan95}). More generally there is a series of solutions at low density where the ion temperature is much higher than the electron temperature. Such a two temperature flow can only be maintained below a critical density (hence mass accretion rate/luminosity) as more frequent collisions give efficient thermalisation between electrons and ions. The specific ADAF solution exists up to a maximum of $\mdot \sim 0.01-0.02$ \citep{Yuan14}, which is where we generally see the state-transitions.

If we scale these transitions (or differing accretion states) up to AGN we expect to see strong UV/EUV emission above $\mdot \sim 0.02$ where the flow can be characterised by a disc-like solution. However, below $\mdot \sim 0.01$, we expect little EUV emission, as now the flow would instead be characterised by a hot X-ray plasma giving a much harder Comptonised spectrum peaking in the hard X-rays rather than the EUV. In this model, the sharp drop in broad-line AGN below $\mdot \sim 0.01$ (e.g. \citealt{Steinhardt10})
is an intrinsic feature rather than a selection effect, as the lack of a strong EUV continuum would naturally lead to an insufficient number of ionising photons required to induce the broad emission lines.

The lack of a UV excess (i.e a "big blue bump") has been seen in nearby low-luminosity AGN \citep{Ho99, Ho08, Nemmen14}, and in limited samples at higher redshifts \citep{Trump11}. More recently, a few, very rare, changing-state (or changing-look) AGN exhibit behaviour where the disappearance of their broad emission lines coincides with a dramatic drop in the EUV continuum \cite{Noda18}, indicative of a transition in accretion state. 
\rev{This is now being strengthened, with recent samples of changing-look AGN showing they tend to reside around this transition region $\mdot \sim 0.01$ \citep{Panda24}.} 
\rev{In addition, recent tidal disruption events appear to support a change in accretion structure dependent on the mass-accretion rate \citep[e.g][]{Wevers21, Yao22, Sfaradi22}}
However, it has not been established whether this transition is present across the wider AGN population \citep{Maoz07} and over cosmic time.

In this paper we use X-ray emission to identify AGN, as (unlike BHBs) they always show a significant high energy tail \citep{Elvis94, Lusso16}, and so should allow us to select objects both below and above the $\mdot \sim 0.01$ transition (e.g \citealt{Aird18}). We use the eROSITA eFEDS X-ray data \citep{Liu22} combined with optical imaging of the eFEDS field using Hyper Suprime-Cam (HSC) to select X-ray sources that are co-incident with galactic nuclei \citep{Li24}, and whose X-rays are unabsorbed. Importantly, the excellent HSC multi-band (\emph{grizy}) imaging, provided by the Subaru Strategic Program \citep{Aihara22} allows for detailed decomposition of the unresolved AGN emission from the host galaxy \citep{Ishino2020, Li21a, Li24}, allowing us to confidently reconstruct optical emission from low luminosity AGN where the total emission is generally dominated by the host galaxy. We then stack the data from each source in bins of black hole mass and monochromatic $3500$\,\AA\, luminosity (used as a proxy for $\mdot$), to obtain mean AGN SEDs within each bin. This allows us to confidently model the emission both below and above the $\mdot \sim 0.01$ transition, showing clearly a change in the accretion structure from a flow dominated by a disc-like continuum to one dominated by an X-ray plasma, as $\mdot$ decreases.

The paper is organised as follows. In section \ref{sec:sample} we provide an overview of the sample used in this paper. Then in section \ref{sec:mods} we give an oveview of our SED model, and apply it to the mean SEDs within each bin, showing conclusively the transition in accretion state. Finally in section \ref{sec:conc} we give a discussion on our results along with our conclusions. Throughout we will assume a standard cosmology, from the Planck 2018 results \citep{Planck20}.

\section{The Sample}
\label{sec:sample}

\subsection{Sample Definition}

We base our study on the combined HSC-eROSITA eFEDS AGN sample described in \citet{Li24} (HSC) and \citet{Liu22} (eROSITA). Briefly, the eFEDS region has 22079 AGN identified from X-ray point sources with ancillary optical-mid IR data to allow photometric redshift estimations \citep{Salvato22}. Our study requires a confident decomposition of the AGN from the host galaxy. This decomposition is most reliable in the redshift range $0.2 \leq z \leq 0.8$, and restricting the sample to this range results in 4975 X-ray AGN. These are cross-matched to the HSC catalogue using the optical counter-part positions from \citet{Salvato22}, giving 3796 X-ray AGN with excellent imaging data for confident host-AGN flux decomposition, described in \citet{Li21a, Ding21}. As our study is focused on the nuclear emission originating from the AGN itself we select only sources that are X-ray unobscured ($N_{H} < 10^{22}$\,cm$^{-2}$, using the values measured in \citet{Liu22}), which reduces the sample to 3509 AGN \citep{Li24}. Additionally, to ensure sufficient signal-to-noise in the X-ray detection we discard any sources with less than 10 counts, $N_{\gamma}$ (as seen by eROSITA).

The HSC decomposition separates the unresolved nuclear emission, modelled by the HSC PSF, from the extended stellar emission from the host galaxy
modelled with a smooth analytic Sersic function using the 2D image decomposition software tool {\sc lenstronomy} \citep{Birrer18, Ding21}, 
(see: \citealt{Ding21, Li21a}). For additional details on the decomposition methodology for our sample we refer the reader to \citet{Li24}. We also refer the reader to Appendix \ref{app:hsc_decomp} for details showing the effect of the decomposition on the SED.

In order to confidently asses the AGN flux for the low-luminosity AGN we only include sources where the AGN optical flux is \emph{at least} $5\,\%$ of the host galaxy flux in all HSC bands, since below this the decomposition results become unreliable. Additionally we also discard sources with a poor fit to the HSC image (defined as a source with $\chi^{2}_{\nu} > 5$). 

In summary, our selection critera can be written as follows:

\begin{itemize}
    \item $F_{\rm{AGN, HSC}} \geq 0.05 F_{\rm{Host, HSC}}$
    \item HSC Image $\chi^{2}_{\nu} < 5$
%    \item $F_{\rm{HSC}} \neq \rm{NaN}$
%    \item $F_{\rm{HSC}} > 0$
    \item $N_{H, \rm{X-ray}} < 10^{22}$\,cm$^{-2}$
    \item $N_{\gamma, \rm{X-ray}} \geq 10$
\end{itemize}

These selection criteria reduce our final sample to 2759 sources. 

\subsection{Mass Estimates}

Only a small portion of our sample (275 sources) overlap with the SDSS DR 16 AGN catalogue, which provides mass estimates from broad-emission lines \citep{Wu22}. This is not necessarily surprising as we have intentionally used an X-ray selection that picks AGN regardless of the presence of a blue-continuum and/or broad emission lines 
such that we can test whether low-luminosity AGN have these features or whether they 
look like the changing-state AGN \citep{Noda18}.
However, we do still require black hole mass estimates for all our sources in order to estimate their $\mdot$. 
The HSC images have stellar mass estimates for the host galaxy (see \citealt{Li24}), and so we use these to estimate the black hole mass from the local $M_{*} - \Mbh$ relation \citep{Ding20}:

\begin{equation}
    \log_{10} \left(\frac{M_{\rm{BH, local}}}{10^{7}} \right) =
    0.27 + 0.98 \log_{10} \left( \frac{M_{*}}{10^{10}} \right)
    \label{eqn:Mloc_ding_etal}
\end{equation}

There is a redshift dependent offset to this relation, as shown in \citet{Li21b}, which we add to Eqn. (\ref{eqn:Mloc_ding_etal}) to give the following $M_{*} - \Mbh$ relation:

\begin{equation}
    \log_{10} \Mbh = \log_{10} M_{\rm{BH, local}} (M_{*}) + \Delta \Mbh(z)
\end{equation}

where

\begin{equation*}
    \Delta \Mbh(z) = \gamma \log_{10} (1+z) \,\,\,\, \rm{where} \,\, \gamma = 1\rev{.00} \pm 0.07
\end{equation*}

\begin{figure}
    \centering
    \includegraphics[width=\columnwidth]{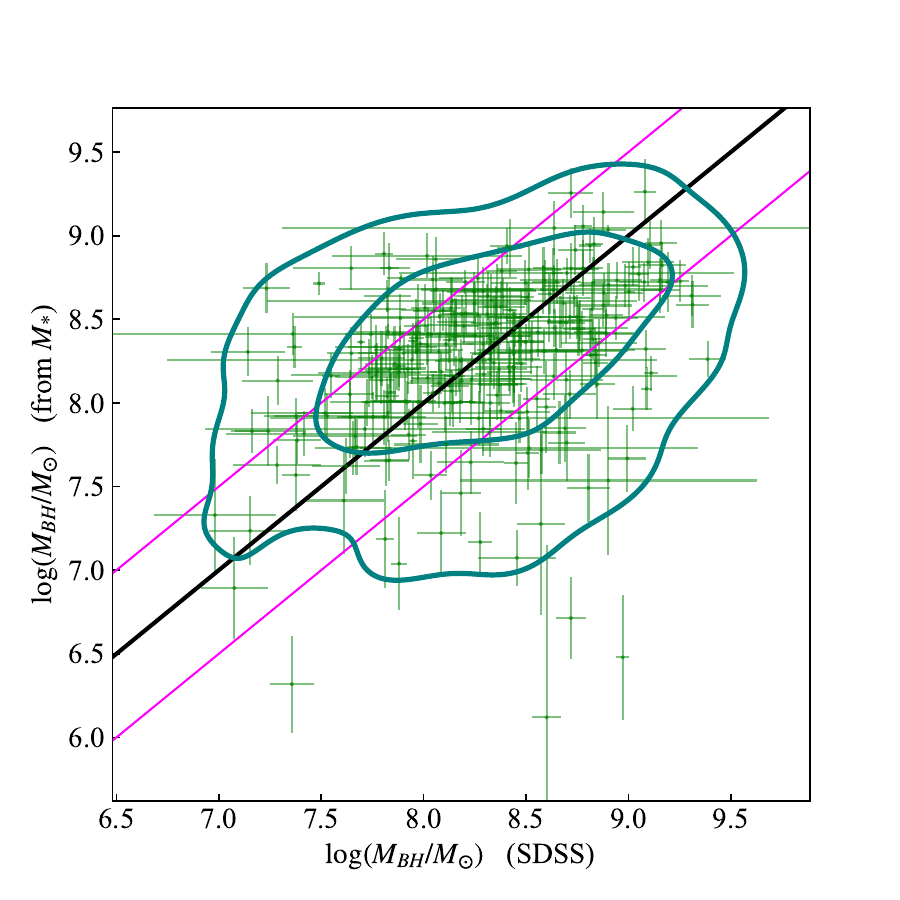}
    \caption{Black hole masses derived from the local $M_{*}$ relation \citep{Ding20, Li21b} versus masses measured from single epoch SDSS spectra in the DR16 AGN catalogue \citep{Wu22}. The solid black line shows a 1:1 relation, while the magenta lines show a $\pm 0.5$\,dex offset. The contours highlight the density distribution of the points, showing the $1\sigma$ and $2\sigma$ levels (inner and outer contours respectively).}
    \label{fig:Mstel_v_Msdss}
\end{figure}

As a sanity check we compare our derived black hole masses to the SDSS measured masses, for the 275 sources where we have overlap. This is shown in Fig. \ref{fig:Mstel_v_Msdss}. It is clear from this that our mass-estimates work as an acceptable proxy, with the majority lying within 0.5\,dex of the SDSS measured values. Additionally, it is also clear that our mass estimates are most reliable for black hole masses in the range $\log_{10} \mmsun = 8.0 - 8.5$, which is where we will focus the main analysis as this is the range where we also have the largest number of sources (see section \ref{sec:m_l_dist}).

\subsection{Calculating Rest Frame Luminosities}

Our sample covers a range in redshift (0.2-0.8), and so we need to convert all the observed fluxes to rest-frame luminosities on a common wavelength/energy grid before we can stack to create mean SEDs.
Here we give a brief description of our methodology, for both the HSC and eROSITA data.

\subsubsection{HSC}

The HSC filters range from the g-band ($\sim 4000$\,\AA$ - 5500$\,\AA) to the Y-band ($\sim 9000$\,\AA$ - 11000$\,\AA). For the redshift range of our sample this corresponds to a rest frame wavelength range of $\sim 3300$\,\AA$ - 6000$\,\AA, to ensure all of our sources fall within this range. We define common rest-frame wavelength bins from 3350\,\AA\, to 5550\,\AA\,, each with a bin width of $\Delta \lambda = 550$\,\AA\rev{, as this gives evenly spaced bins \revv{that have a width close to the HSC filters yet are narrow enough to show changes in spectral shape}}.

For each source we first de-redden the observed frame AGN flux, using the extinction law of \citet{Fitzpatrick07}, and $E(B-V)$ values from the dust maps of \citet{Schlafly11}\rev{, stressing that this only corrects for the extinction within our own Galaxy}. \revv{No attempt has been made to correct for extinction/reddening in the frame of the AGN.}. \revv{We also include a cosmological K-correction on the flux-densities, by multiplying with a factor $(1+z)^{-1}$}. We then convert the observed wavelength bins to rest frame wavelength; $\lambda_{em} = \lambda_{obs}/(1+z)$. For each of the common rest-frame bins, $i$, we then calculate the fractional overlap, $w$, with each of the HSC rest-frame bins, $j$. This is defined such that $w_{j}=1$ implies that the 
%original bin is fully contained within the common rest-frame bin \rev{or the 
\rev{common rest-frame bin is fully contained in the HSC rest-frame bin (since we are working in flux-density units)}, $w_{j} < 1$ implies the bins only partially overlap, and $w_{j} = 0$ implies no overlap between the common and original bin. These are then used to calculate the flux-density in the new common rest-frame bin, $F_{c, i}$, using a weighted log-mean:

\begin{equation}
    \log_{10} F_{c, i} = \frac{\sum_{j} w_{j} \log_{10} F_{\rm{HSC}, j}}{\sum_{j} w_{j}}
\end{equation}
where $F_{\rm{HSC}, j}$ is the HSC flux-density in the filter $j$ (where $j$ iterates over all HSC filters). We note that this does approximate the sensitivity of each HSC filter as a top-hat, but this should be 
adequate as the transmittance of each HSC filter is mostly uniform over their respective range \citep{Satoshi18}.  Any error introduced will be significantly smaller than the dispersion in the stacked SEDs (see Fig. \ref{fig:SED_fig})

We stress here that the HSC flux-densities used are the \emph{decomposed} AGN fluxes from \citet{Li24}, so do not contain host-galaxy contamination. 
These flux-densities are then simply converted to luminosity densities through the luminosity distance, $d_{L}$, of each source, calculated from their redshift using our assumed cosmology \citep{Planck20}.

\subsubsection{eROSITA}

%The eROSITA data are slightly more complicated, as they need to be corrected for line-of-sight absorption within our galaxy (we stress that this is \emph{extrinsic} absorption, not \emph{intrinsic} absorption associated with the AGN itself). For this we use {\sc xspec} v.12.13.0c \citep{Arnaud96}.

The eROSITA data need to be unfolded from the instrument response before we can re-bin them onto a common grid. For this we use {\sc xspec} v.12.13.0c \citep{Arnaud85}

For each eROSITA spectrum we apply a simple absorbed power-law model using {\sc tbabs*pow}, where the galactic column in the absorption model {\sc tbabs} is fixed at that from  \citet{HI4PI16}, and {\sc pow} gives the power-law component, where we use the values derived by \citet{Liu22}. We stress that unlike \citet{Liu22}, we \emph{do not} include an intrinsic absorption component to model any potential absorption from within the AGN itself \rev{(i.e they used {\sc tbabs*ztbabs*pow}, while we omit the {\sc ztbabs} component)}. \rev{We neglect the intrinsic absorption component because our sample, by design, should only include unobscured sources, and hence we assume the observed spectral shape after correcting for Galactic absorption is representative of the true SED. This assumption is tested in detail in Appendix \ref{app:Nh_test}}
%This is in part as we have already discarded sources showing significant intrinsic absorption, but mostly because we wish to ensure that our analysis is in fact showing an intrinsic change in the AGN SED, and not an effect induced by spurious absorption.

We use the above model to unfold the raw data from the instrument response, \rev{noting that it makes no assumptions on any spectral complexity beyond a power-law,} providing raw flux densities. We then remove the absorption component from the model, and write out the unabsorbed model. The energy dependent absorption correction factor for each eROSITA energy bin is simply the ratio of the unabsorbed to absorbed model, which is then applied to the raw flux-densities to give the galactic absorption corrected flux-densities.

The absorption corrected flux densities 
are then corrected for redshift and re-binned onto a common energy grid. eROSITA extends from 0.2\,keV to 8\,keV. Since the sensitivity of eROSITA drops off rapidly at higher energies \citep{Predehl21}, we choose to limit the upper observed frame energy to 5.3\,keV. For our redshift range this gives rest-frame energy limits from 0.36\,keV to 9.6\,keV. We use these to define a common rest-frame energy grid, with bin edges at [0.4, 0.7, 1, 2, 4, 8]\,keV. These bins are intentionally wider than the intrinsic eROSITA bins in order to simplify the rebinning process and maximise signal-to-noise by providing higher photon counts per bin. First we find all original energy bins (shifted to the AGN rest-frame) that lie within the new energy bin. We then calculate the log-mean flux density of all original bins that lie within the new bin, using the absorption corrected flux-densities. This then provides the flux-density within the new energy bin.

As with the HSC flux-densities, these are converted to luminosity densities through the luminosity distance, again using the source redshift and Planck 2018 Cosmology results \citep{Planck20}.

\subsection{Redshift Distribution}
\label{sec:red_dist}

%\begin{figure*}
%    \centering
%    \includegraphics[width=0.8\textwidth]{Figures/Lopt_v_redshift-crop.pdf}
%    \caption{Redshift and 3500\,\AA\, monochromatic luminosity distribution across the ({\it left}) full sample and ({\it right}) within our fiducial $\log_{10} \mmsun \in [8.0, 8.5]$ mass-bin. The top histograms show the redshift distribution of the samples, normalised such that their integral is unity. It is clear that the distributions do not change significantly between the full sample and within our fiducial mass-bin. There is a slight deficit in the number of sources below $z \sim 0.25$, however the remaining redshift range (up to $z=0.8$) has a relatively even distribution of sources.}
%    \label{fig:z_Lopt}
%\end{figure*}

%\begin{figure*}
%    \centering
%    \includegraphics[width=0.8\textwidth]{Figures/Lx_v_redshift-crop.pdf}
%    \caption{As in Fig. \ref{fig:z_Lopt}, but for 2\,keV luminosity.}
%    \label{fig:z_Lx}
%\end{figure*}

\begin{figure*}
    \centering
    \includegraphics[width=0.9\textwidth]{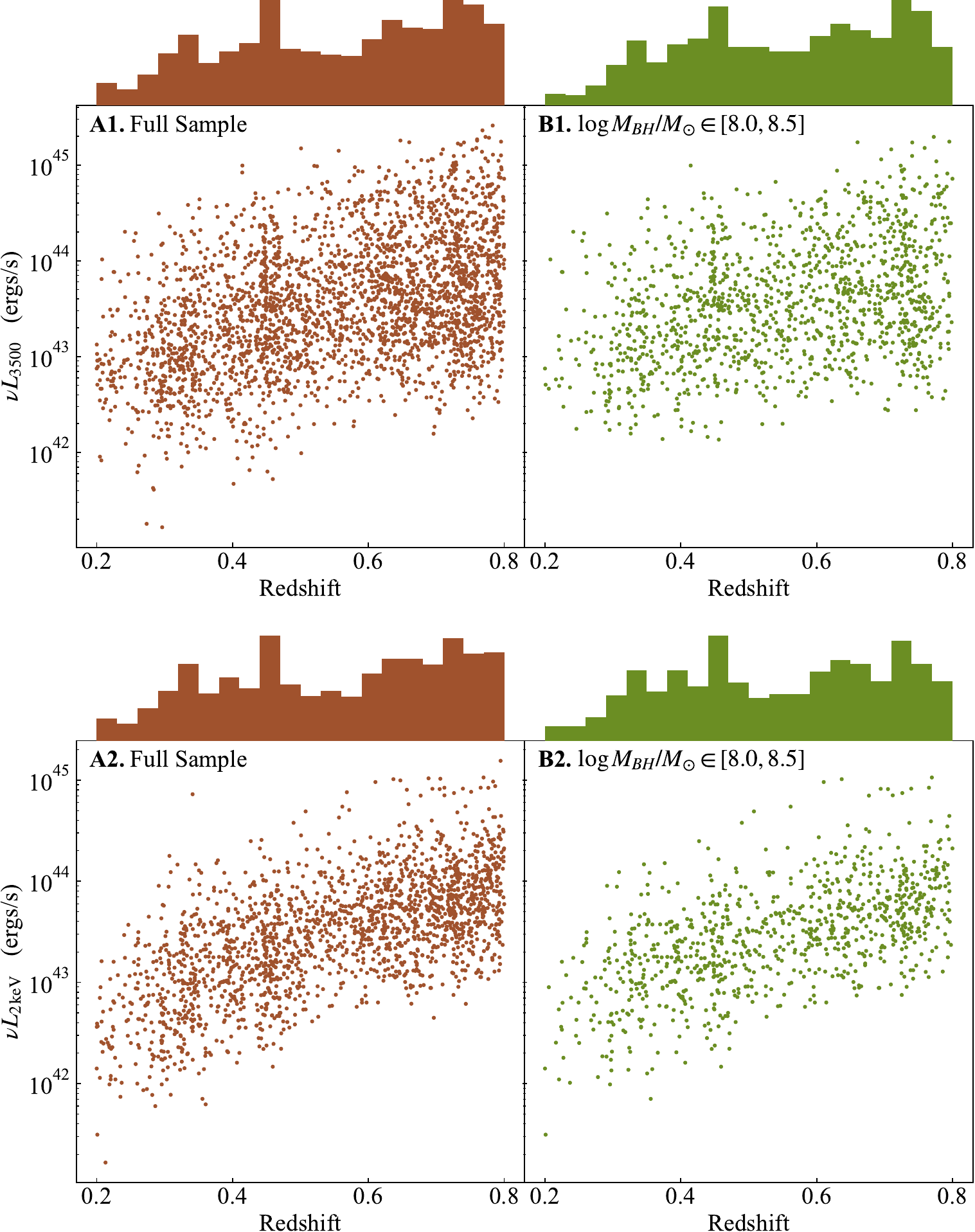}
    \caption{Redshift and monochromatic luminosity distributions for 3500\,\AA\, (\emph{top}) and 2\,keV (\emph{bottom}). In both cases the left panel shows the full sample and the right the sub-sample within our fiducial $\log_{10} \mmsun \in [8.0, 8.5]$ mass-bin. The top histograms show the redshift distribution of the samples, normalised such that their integral is unity. It is clear that the distributions do not change significantly between the full sample and our fiducial mass-bin. In all cases there is a slight deficit in the number of sources below $z \sim 0.25$, however the remaining redshift range (up to $z = 0.8$) has a relatively even distribution.}
    \label{fig:z_L}
\end{figure*}

The left panels of Figure \ref{fig:z_L} show the redshift distribution of our sample for 3500\,\AA\, monochromatic luminosity (\emph{top}) and 2\,keV monochromatic luminosity (\emph{bottom}) respectively, using the redshift measured in \citet{Salvato22}. The right panel of each figure shows the 
distribution for the subset of sources, with black hole masses within $\log_{10} \mmsun \in [8.0, 8.5]$, which we use for further analysis (see section \ref{sec:m_l_dist}).

The source numbers (top histogram) are relatively evenly distributed with redshift in both 3500\,\AA\, luminosity and 2\,keV luminosity in both the full sample and the $\log_{10} \mmsun \in [8.0, 8.5]$ mass bin apart from a drop in numbers below $z\le 0.25$ due to the small volume. Similarly, the mean source luminosity increases with redshift as the larger volume includes the rarer, more luminous objects. 

\subsection{Mass and Luminosity Distribution}
\label{sec:m_l_dist}

\begin{figure*}
    \centering
    \includegraphics[width=0.7\textwidth]{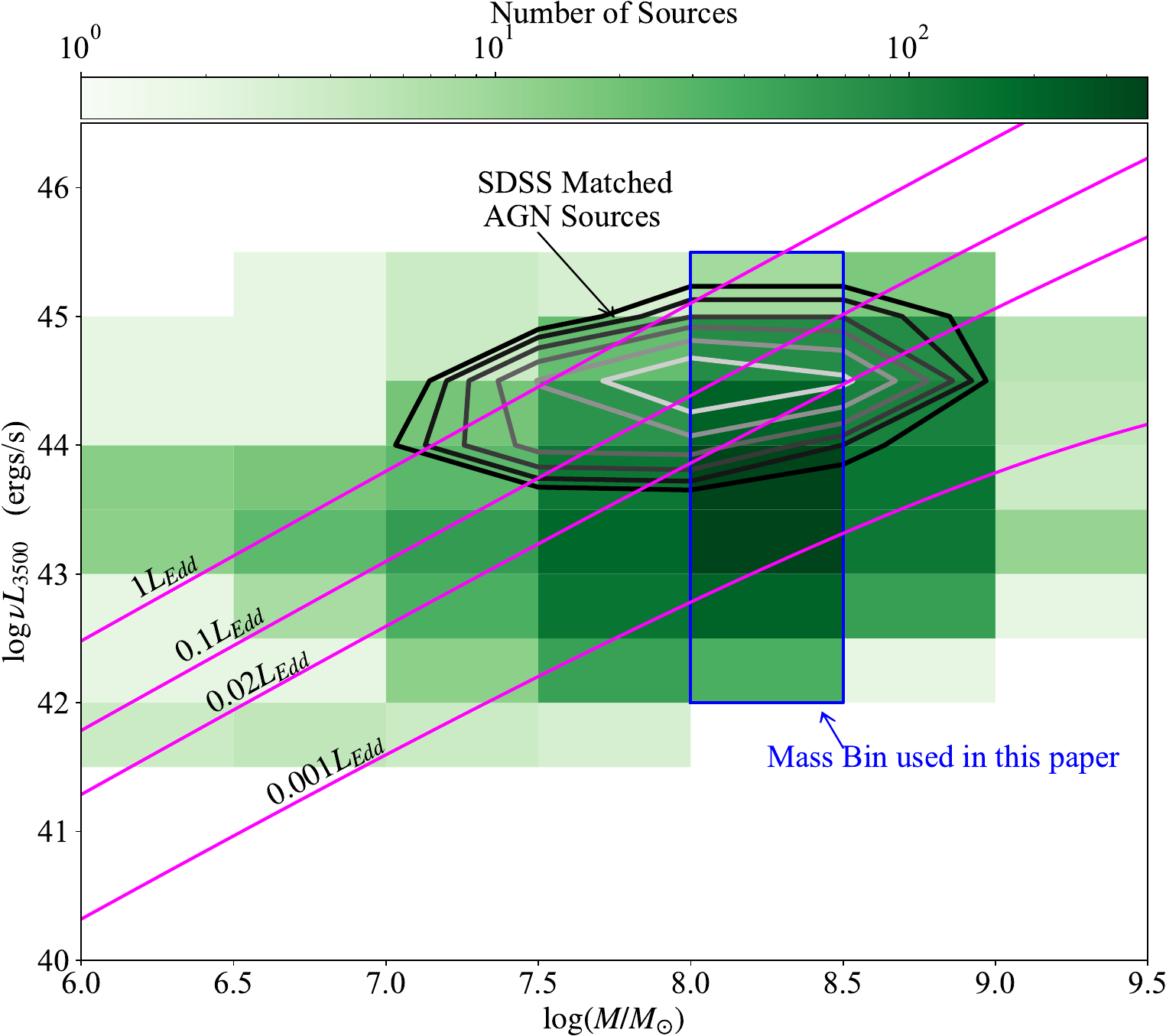}
    \caption{Distribution of our sample across bins of black hole mass and monochromatic $3500$\,\AA\, luminosity. The green shading gives the number of sources per bin (as defined by the colour-bar on top). The black contour lines show the distribution of sources that have matches with the SDSS DR16 AGN catalogue \citep{Wu22}. Magenta lines show the predicted $3500$\,\AA\, luminosity at different Eddington ratios, calculated for a standard accretion disc extending to the ISCO, while the blue box indicates the bins used for the main analysis presented in this paper. It is clear that our eROSITA-HSC combined sample pushes to significantly lower optical luminosity than the broad-line SDSS sources, allowing us to probe AGN accretion below the $2\,\%$ of Eddington transition.}
    \label{fig:lM_lL_dist}
\end{figure*}

\begin{figure}
    \centering
    \includegraphics[width=\columnwidth]{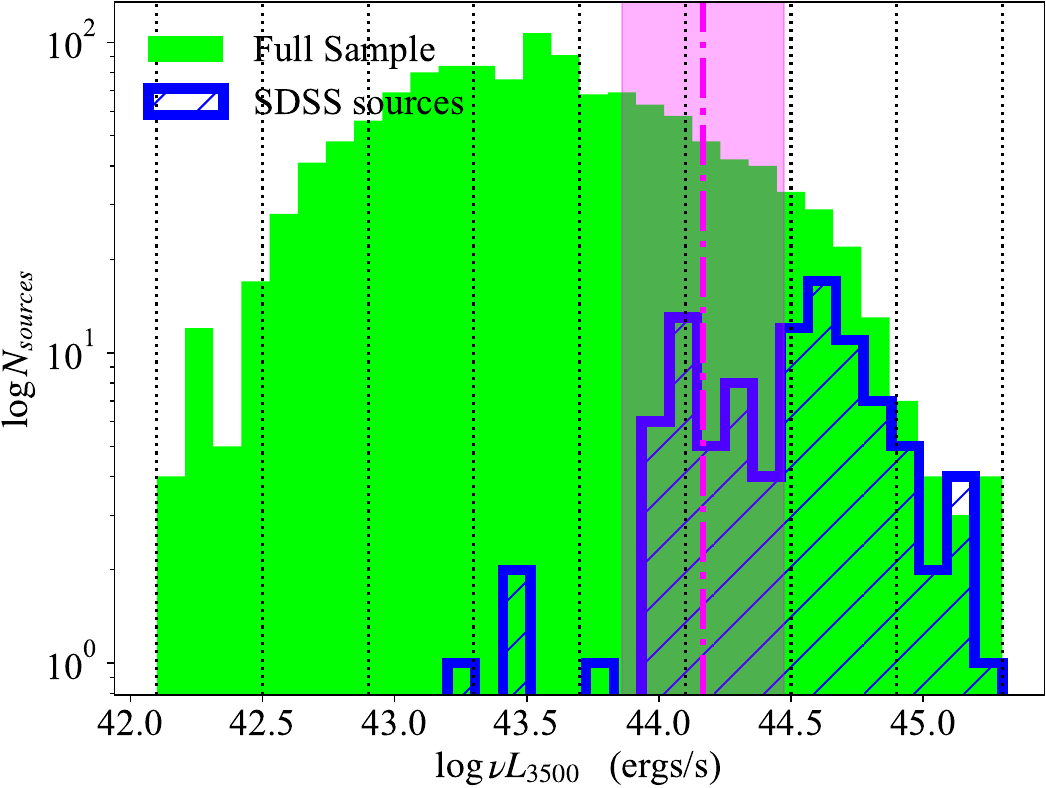}
    \caption{The $3500$\,\AA\, luminosity distribution for the sources within the $\log_{10} \mmsun \in [8.0, 8.5]$ mass bin. The green bars show the distribution for the full sample, while the blue-dashed bars show the distribution of the sources in the SDSS DR16 AGN catalogue \citep{Wu22}. The shaded magenta region band shows the $3500$\,\AA\, monochromatic luminosity at $2\,\%$ of Eddington for the black hole mass range $\log_{10} \mmsun \in [8.0, 8.5]$, calculated for a standard disc extending to the ISCO. The dashed-dotted magenta line shows this for the geometric centre of the mass-bin. \rev{The vertical black dotted lines show the luminosity binning used in the SED analysis in this paper}. It is clear that not only does our X-ray selected sample extend to considerably lower optical luminosity than the SDSS matched sample, but it also peaks below $2\,\%$ of Eddington, unlike the SDSS matched sample.}
    \label{fig:L3500dist}
\end{figure}

Each AGN within our sample now has a rebinned, deabsorbed rest frame spectrum with a black hole mass estimate. We start by subdividing our sample into black hole mass bins, from $\log_{10} \mmsun = 4$ to $\log_{10} \mmsun = 9.5$ with bin width 0.5\,dex \rev{(note that we only show bins with $\log \mmsun \geq 6$ in Fig. \ref{fig:lM_lL_dist} due to limited sources below this)}. This allows us to control for changes in the SED due to differences in black hole mass leading to intrinsic differences in the temperature of the accretion flow (i.e a larger black hole gives a cooler accretion disc; \citealt{Shakura73}). We have chosen relatively broad bins of 0.5\,dex to allow for the intrinsic uncertainty in the black hole mass estimates. Fig. \ref{fig:Mstel_v_Msdss} show that the majority of our mass estimates (from the host galaxy stellar mass) lie within 0.5\,dex of the single epoch mass estimates from SDSS DR 16 \citep{Wu22} (which itself has an intrinsic uncertainty). Hence we do expect there to be some leakage between mass bins, however by choosing broad bins the overall effect of this on the stacked SEDs should be minimal.

%Hence, the sample is subdivided into 3500\,\AA\, luminosity bins from $\log_{10} \nu L_{3500} = 42$\,ergs/s to $\log_{10} \nu L_{3500} = 46$\,ergs/s, again with bin widths of 0.5\,dex.

%Overall this leads to the sample distributed over a 2 dimensional grid, shown in Fig. \ref{fig:lM_lL_dist}. 

Within each mass-bin we further subdivide the sample into bins of monochromatic optical luminosity, as this should originate from the disc and as such trace the mass-accretion rate, which significantly impacts the shape of the SED \citep{Kubota18}. We choose the monochromatic luminosity at 3500\,\AA, as the bluer the wavelength the closer to the SED peak it will originate from, providing a better tracer of mass-accretion rate. \rev{This leads to a 2D-distribution, visualised in Fig. \ref{fig:lM_lL_dist}}
Here each mass-luminosity grid is shown as a green rectangle, where the shade indicates the number of sources within each bin (with a darker green indicating more sources, as shown by the colour-bar). 
\rev{We note that though a regular grid is used for to visualise to overall sample, in the SED analysis later we optimise the luminosity bin edges for each mass-bin separately to give contiguous bins with an even distribution throughout each luminosity bin (in order to avoid side-weighted distributions in the edge bins).} 
The black-grey contours show the distribution of our sources that are also included in the SDSS DR16 AGN catalogue \citep{Wu22}. These are objects selected for spectroscopic followup based on their photometric colours showing a blue continuum \citep{Richards02}, and these spectra show the characteristic broad emission lines. This immediately shows that our X-ray selected sample extends down to considerably lower optical AGN luminosity than that of the SDSS AGN catalogue. This is more clear in Fig. \ref{fig:L3500dist}, which shows the 3500\,\AA\, luminosity distribution for a single mass bin, $\log_{10} \mmsun \in [8.0, 8.5]$, for both the full sample (green, solid) and the sources matched to the SDSS AGN catalogue (blue, hatched). Here it is clear that not only does our sample extend to considerably lower optical luminosity, it also peaks at lower luminosity, indicating that bright broad-line AGN are not necessarily the dominant unobscured AGN population.

A more interesting metric than the monochromatic luminosity is the Eddington ratio. This is shown by the magenta lines in Fig. \ref{fig:lM_lL_dist}. We stress here that these are calculating from converting the 3500\,\AA\, luminosity to bolometric luminosity via a standard \citet{Shakura73} accretion disc model, and as such the precise values should be treated with caution. For $\LLedd \lesssim 0.02$, where we see the changing-look transition, we expect a vastly different accretion structure \citep{Done07, Noda18}, and so converting via a disc model is not accurate. However, it is sufficient to give an estimate of what distribution of sources lie above or below the $2\,\%$\,Eddington transition. It is clear that the majority of the SDSS AGN sources lie above this transition, as seen previously in \citet{Mitchell23}, while the majority of our sample lie below (for black hole masses $\log_{10} \mmsun \gtrsim 7$ - below this we do not trust the mass estimates). Again this becomes more evident in Fig. \ref{fig:L3500dist}, where we see that our sample also peaks below the $2\,\%$ transition. This already gives strong evidence for a change in SED at low Eddington ratios, where the optical luminosity drops much faster than the X-ray luminosity, as predicted if there is a transition in the physical nature of the accretion flow from the optical being dominated by a disc to an ADAF.

We investigate this in more detail by stacking the spectra within each bin, averaging in logarithmic luminosity on the common wavelength/energy binning. This gives an average SED in each mass-luminosity bin. For the remainder of the paper we focus on the single mass bin $\log_{10} \mmsun \in [8.0, 8.5]$, as this is the bin where the black hole mass estimates from the host galaxy are most reliable, and also where we have most objects. \rev{We note here that the following analysis is also performed for the bins $\log_{10} \mmsun \in [7.5, 8.0]$ and $[8.5, 9.0]$ as a consistency check. These are presented in Appendix \ref{app:mbin})}. \rev{Using a single mass bin} gives a sequence of SEDs for changing monochromatic 3500\,\AA\, luminosity at a fixed mass, shown in Fig. \ref{fig:SED_fig} (orange (HSC) and blue (eROSITA) points), 
\rev{where the luminosity bins are chosen to precisely fit around the underlying luminosity distribution, giving bins of width 0.4\,dex, each with an even luminosity distribution within, as shown by the vertical black dotted lines in Fig. \ref{fig:L3500dist}}. 
It is immediately clear that the 3500\,\AA\, monochromatic luminosity changes by over 3 orders of magnitude while the X-ray power changes by less than a factor of a few. It is also immediately clear that the shape of the optical spectrum changes dramatically, from a blue continuum like a standard disc in the higher luminosity bins, to a fairly flat continuum in the middle bins, to a strongly red continuum very unlike a standard disc in the lowest luminosity bins. Again, this suggests a dramatic change in the accretion structure, which we explore in greater detail in the next section (\ref{sec:mods}), where we detail our SED model and apply it to the stacked SEDs to give a physical interpretation of the change in SED shape.

\section{Modelling the SEDs}
\label{sec:mods}

To explore the spectral evolution in detail we fit the stacked AGN SEDs within our fiducial mass bin of $\log_{10} \mmsun \in [8.0, 8.5]$. We do not use standard \citet{Shakura73} disc models, as these have multiple well known issues. Firstly, there is the high energy X-ray tail which is always present at a significant level \citep{Elvis94, Lusso16}. More seriously for the disc models themselves, the UV spectrum often shows a downturn before the predicted peak, which appears to connect across the data-gap (due to ISM absorption, e.g \citealt{Laor97, Telfer02}) to an upturn in the soft X-ray bandpass (the soft X-ray excess) (e.g \citealt{Czerny87, Laor97, Mehdipour11, Jin12a, Petrucci18, Kubota18}). The standard disc models also predict that the time-scales for intrinsic variability is extremely slow (thousands of years), in conflict with multiple observations (e.g \citealt{Noda18, Stern18, Hernandez20, Neustadt22, Yao23, Neustadt24}). Nonetheless, the ubiquity of the strong UV emission clearly indicates that there is optically thick material emitting a large fraction of the gravitational potential energy from accretion, i.e there is a disc but its properties (spectrum and variability) are different to those predicted by standard models.

Hence we use modified disc models, which assume that the disc vertical structure can be different than in standard models such that it does not completely thermalise \citep{Rozanska15, Jiang20, Kawanaka24}. This could then form a warm Comptonising plasma, which is optically thick, giving rise to a Comptonised disc-spectrum, rather than a pure multi-colour black-body disc spectrum; naturally explaining the UV turnover linking into the soft X-ray excess \citep{Mehdipour11, Done12, Mehdipour15, Kubota18, Petrucci18}, and generally comprises the dominant component of the disc emission \citep{Czerny87, Magdziarz98, Done12, Petrucci13, Petrucci18}. However, the model does still allow the emission to revert back to a standard black-body disc at larger radii \citep{Done12, Jin12a, Kubota18}.

In addition to a disc and disc-like structure, the model allows for the accretion flow to evaporate into an optically thin hot X-ray plasma in the innermost regions, $r_{\rm{isco}} < r < r_{\rm{hot}}$, giving rise to the Comptonised high energy X-ray tail.

Hence, our accretion structure consist of three distinct regions: An inner X-ray plasma for $r_{\rm{isco}} < r < r_{\rm{hot}}$, which then condenses into a warm optically thick disc-like structure for $r_{\rm{hot}} < r < r_{\rm{warm}}$, and then eventually reaching thermalisation forming a standard disc for $r > r_{\rm{warm}}$ \citep{Kubota18}. These models generally give a good fit to the observed SEDs of individual AGN with significantly higher quality/broader bandpass data (e.g \citealt{Matt14, Matzeu16, Done16, Czerny16, Hagino17, Porquet18, Porquet19, Porquet24}), and comes with the additional feature that the non-standard optically thick region gives the potential for significantly faster intrinsic disc variability \citep{Hagen24}, more in line with current observations. 

In this section we give a brief overview of the physical implementation of the model in section \ref{sec:agnsed}, using the publicly available {\sc xspec} model {\sc agnsed} \citep{Kubota18}, before applying it to the data in \ref{sec:dat_fit}. Throughout we use the standard notation for radii, where $R$ denotes the radius in physical units and $r$ is the dimensionless radius in gravitational units, related through $R = rR_{G}$ where $R_{G} = GM/c^{2}$.

\subsection{{\sc agnsed} - Physical Implementation}
\label{sec:agnsed}

{\sc agnsed} (see \citealt{Kubota18} for full details) assumes a standard (relativistic) disc \citet{Novikov73} emissivity throughout the entire flow, with a radial temperature profile given by $T_{NT}^{4}(R) \propto R^{-3} f(R)$ where $f(R)$ describes the radial disc structure in the Kerr metric \citep{Page74}.

In the outermost regions of the flow, $r > r_{\rm{warm}}$, this model assumes the flow thermalises to a standard accretion disc. Hence, splitting the flow up into radial annuli of width $\Delta R$, the emission from each annulus will be a standard black-body $B_{\nu}(T_{NT}(R))$ with total luminosity $2 \times 2\pi R \Delta R \sigma T_{NT}^{4}(R)$, where the extra factor $2$ comes from the disc emitting from both sides and $\sigma$ is the Stefan-Boltzmann constant.

In the warm disc-like region, $r_{\rm{hot}} < r < r_{\rm{warm}}$, the vertical disc structure is assumed to change, such that the dissipating region moves up into the photoshpere forming a slab geometry above an underlying passive disc structure. Here the seed photons are tied to the underlying disc, such that their temperature is simply $T_{NT}$ and each annulus has luminosity $4\pi R \Delta R \sigma T_{NT}^{4}(R)$. Additionally, the seed photons are assumed to form a black-body distribution, which then Compton scatters through the photosphere forming a Comptonised black-body spectral shape; modelled internally using {\sc nthcomp} \citep{Zdziarski96, Zycki99}. It is assumed that the disc photosphere is optically thick, and has a covering fraction of unity such that all seed-photons are Comptonised \citep{Petrucci18}. Within {\sc agnsed} the warm Comptonisation region is assumed to have a single electron temperature, $kT_{e, w}$, and photon index, $\Gamma_{w}$, left as free parameters.

{\renewcommand{\arraystretch}{1.6} %Adjust table spacing
\begin{table*}
    \centering
    \begin{tabular}{c | c c c c c c c}
        $\log \nu L_{3500}$ (ergs/s) & $\log \dot{m}$ & $kT_{e, w}$ (keV) & $\Gamma_{h}$ & $\Gamma_{w}$ & $r_{\rm{hot}}$ & $T_{hi}$ (K) & [$r_{w} (T_{hi})$] \\
        \hline
        
        \rev{$[42.1, 42.5]$} & \rev{$-1.93$} & $^{\dagger}$ & \rev{$1.53$} & $^{\dagger}$ & \rev{$306^{\dagger \dagger}$} & \rev{$4611$} & \rev{[$306$]} \\

        \rev{$[42.5, 42.9]$} & \rev{$-2.04$} & $^{\dagger}$ & \rev{$1.65$} & $^{\dagger}$ & \rev{$224^{\dagger \dagger}$} & \rev{$5464$} & \rev{[$224$]} \\

        \rev{$[42.9, 43.3]$} & \rev{$-2.04$} & $^{\dagger}$ & \rev{$1.79$} & $^{\dagger}$ & \rev{$189^{\dagger \dagger}$} & \rev{$6054$} & \rev{[$189$]}\\

        \rev{$[43.3, 43.7]$} & \rev{$-1.93$} & \rev{$0.36$} & \rev{$1.92$} & \rev{$2.60$} & \rev{$115$} & \rev{$5920$} & \rev{[$217$]}\\

        \rev{$[43.7, 44.1]$} & \rev{$-1.80$} & \rev{$0.42$} & \rev{$1.93$} & \rev{$2.72$} & \rev{$52.8$} & \rev{$5920^{*}$} & \rev{[$238$]}\\

        \rev{$[44.1, 44.5]$} & \rev{$-1.40$} & \rev{$0.46$} & \rev{$1.69$} & \rev{$2.63$} & \rev{$23.0$} & \rev{$5920^{*}$} & \rev{[$328$]}\\

        \rev{$[44.5, 44.9]$} & \rev{$-0.98$} & \rev{$0.20$} & \rev{$1.71$} & \rev{$3.82$} & \rev{$15.8$} & \rev{$5920^{*}$} & \rev{[$463$]}\\

        \rev{$[44.9, 45.3]$} & \rev{$-0.61$} & \rev{$0.99$} & \rev{$2.35$} & \rev{$3.70$} & \rev{$12.4$} & \rev{$5920^{*}$} & \rev{[$610$]}
        
    \end{tabular}
    \caption{Fit values for each of the luminosity bins (left column) for the stacked SEDs within the mass bin $\log_{10} \mmsun \in [8.0, 8.5]$. We only show parameters that were kept free during the fitting process. The $r_{w}$ values in square brackets show the radius where the disc temperature is $T_{hi}$. \rev{Note that the errors on the derived parameters are too large to be meaningful (due to the significant dispersion in the stacks).}\\
    $\dagger$ The warm Compton region is not required in this bin. \\
    $\dagger \dagger$ No warm Compton region is required so $r_{w}=r_h$, and $T_{hi}=T_{bb}(r_h)$. \\
    $*$ In these bins the standard outer disc is subdominant and $T_{hi}$ cannot be well constrained. Hence we freeze it to the last bin where it could make an impact (i.e $[43.3, 43.7]$).} 
    \label{tab:fitpars}
\end{table*}
}

Below a radius $r_{\rm{hot}}$ the warm disc-like structure evaporates into a hot, optically thin, geometrically thick X-ray plasma; responsible for forming the high energy X-ray tail \citep{Narayan95, Liu99, Rozanska00b}. This plasma is powered by the energy dissipated in the flow between $r_{\rm{hot}}$ and $r_{\rm{isco}}$, still assuming standard \citet{Novikov73} emissivity, giving $L_{\rm{diss}}$. In addition there is a contribution to the total emission from the seed-photons entering the corona from the disc, $L_{\rm{seed}}$, such that the total power emitted by the corona is $L_{\rm{hot}} = L_{\rm{diss}} + L_{\rm{seed}}$ (see \citealt{Kubota18} for details on calculating thse). As with the warm Comptonisation in the disc-like region, it is assumed that the X-ray plasma has a single electron temperature $kT_{e, h}$ and photon index $\Gamma_{h}$, which are left as free parameters.

The total SED is then simply found by adding together the integrated emission from each component within the flow.

A key feature of {\sc agnsed} is the energy balance. The total available power within the flow is always set by the mass, $M_{BH}$, and Mass-Accretion rate, $\dot{M}$, assumed constant at all radii in the flow. The relative contribution to the total luminosity from each of the three regions within the flow is then set by the integrated emissivity over the radii where each component dominates.
As an example, a flow where the disc-like structure covers the majority of the available emitting area down to $r\sim r_{\rm{isco}}$ will have an SED dominated by the warm Comptonisation, whereas a flow with an extensive X-ray plasma with $r_{\rm{hot}}\gg r_{\rm{isco}}$ will have an SED dominated by the hot Comptonisation component with very little disc. In this sense, fitting the AGN SED gives us an estimate of not only the relative contribution to the total output power from each component, but also the relative size scales of each component. 

An important note is that although {\sc agnsed} takes into account the energy balance, it does not consider the conservation of photon number although Compton scattering does. In some sense this is allowing for some uncertainty in the source of seed photons. The model only considers seed photons originating from the disc regions, which in disc dominated states will certainly be sufficient (e.g \citealt{Malzac09}). However, in states dominated by the X-ray plasma there could be an issue with the spectrum being photon-starved when only considering seed-photons from the disc, similar to hard state BHBs (e.g \citealt{Poutanen18}), and will almost certainly require additional seed photons originating within the flow through, e.g, cyclo-scynchrotron.

\subsection{Fitting the data}
\label{sec:dat_fit}

We fit the stacked data within each bin with the {\sc agnsed} model, using the standard $\chi^{2}$ minimisation routine within {\sc xspec} v.12.13.0c \citep{Arnaud96}. For simplicity we treat the $1 \sigma$ dispersion on the flux within each energy/wavelength bin as the error on the flux during the fitting. For mass-luminosity bins with too few sources to confidently calculate the dispersion ($< 10$, only relevant for the highest luminosity bin in our fiducial mass bin), we assume a $10\,\%$ error on the flux. Due to the large dispersion in the data, especially the X-ray, we stress that these fits are meant to display the rough power-output and evolution of the SED, rather than a detailed determination of model parameters.

The black hole mass is fixed to the geometric centre of the mass-bin, which for the bin considered in this paper ($\log_{10} \mmsun \in [8.0, 8.5]$) is $1.7 \times 10^{8}\,\msun$. We conservatively assume spin zero, where the general-relativistic corrections from ray-tracing (not included in the standard version of {\sc agnsed}) are small (see \citealt{Hagen23b} - certainly smaller than the dispersion within the data). As these are unobscured AGN we assume an on average a small inclination angle, fixing it to $\cos(i) = 0.87$ throughout. Finally, we fix the X-ray coronal scale height to $h_{x} = 10$ \rev{(which sets the fraction of X-ray power reprocessed by the disc - see \citet{Kubota18} for details)}, and the outermost radius to $r_{\rm{out}} = 10^{4}$.

Our initial fits (not tabulated) showed that the peak temperature of the outer black-body disc, $T_{hi} = T_{bb}(r_{w})$ remained approximately constant ($\sim 4000 - 6500$\,K), despite the large changes in $\mdot$. Hence, we slightly modified {\sc agnsed} such that this transition temperature is now an explicit free parameter, replacing $r_{w}$, which is now an implicit parameter calculated internally for a given $\mdot$ and $T_{hi}$.

\begin{figure*}
    \centering
    \includegraphics[width=\textwidth]{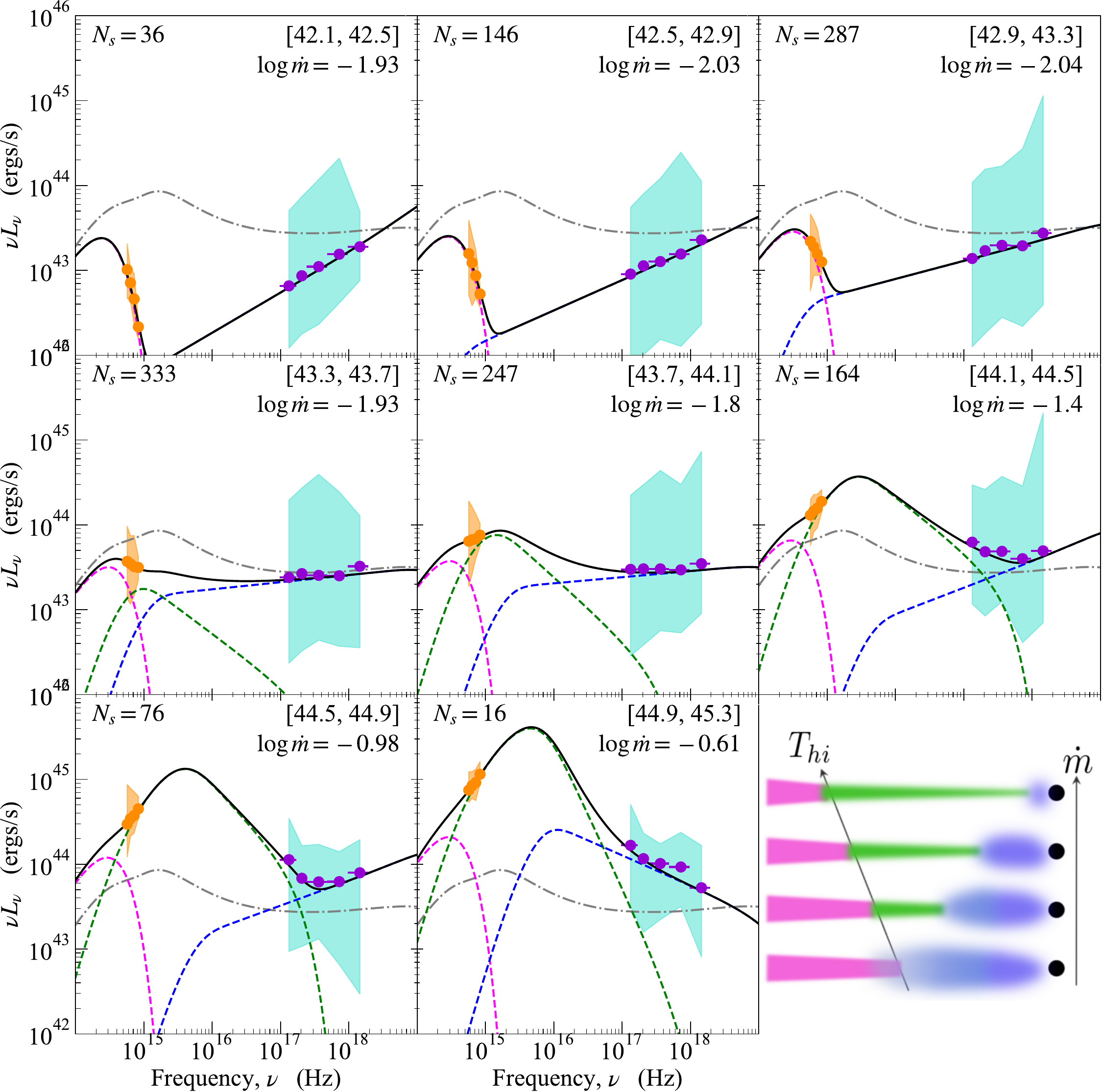}
    \caption{Stacked SEDs for the mass bin $\log_{10} \mmsun \in [8.0, 8.5]$. Each panel shows a single monochromatic, $\nuLnu$, $3500$\,\AA\, luminosity bin, with the precise bin range shown in ergs/s by the bracketed values in the top right corner of each panel. The number of sources, $N_{s}$, within each bin is given in the top left corner. The binned HSC data are shown by the orange points, while the binned eROSITA data are given by the \rev{violet} points. The orange and cyan regions show the dispersion on the HSC and eROSITA data respectively. Each bin has been fit with {\sc agnsed}, with the resulting model SED shown by the solid black line, while the coloured dashed lines show the individual components to the total model. Here magenta indicates the outer standard disc, green the unstable warmly Comptonised disc, and blue the inner X-ray corona. The dashed-dotted grey line shows the model for the middle bin \rev{($\log_{10} \nu L_{3500} \in [43.7, 44.1]$\,ergs/s)}, provided to highlight the relative change in the SED as we move through the luminosity bins. The bottom right panel shows a cartoon of the change in accretion geometry with $\mdot$. The colours in the cartoon corresponds to the colours in the SED components.}
    \label{fig:SED_fig}
\end{figure*}

The resulting best fit parameters are shown in Table \ref{tab:fitpars}, and the resulting SEDs are shown in Fig. \ref{fig:SED_fig}. As well as the (expected) increase in $\mdot$ with optical luminosity, we see a correlated decrease in $r_{\rm{hot}}$. This shows how the inner X-ray plasma/corona becomes less and less dominant as $\mdot$ increases \citep{Lusso16}. This is also seen in the data themselves (Fig. \ref{fig:SED_fig}), where the X-ray emission has comparatively little change (only a factor of a few) whereas the optical luminosity (and hence $\mdot$) change by multiple orders of magnitude. Our model relies on the energy balance to determine the size-scale of the components, so the approximate constant X-ray power (at $L_{\rm{hot}} \sim 0.02 L_{\rm{Edd}}$ - \citealt{Kubota18}) means that it requires a smaller fraction of the total accretion power as $\mdot$ increases, and so $r_{\rm{hot}}$ decreases. The increasing dominance of the disc means means there are more seed photons cooling the hot corona, and so $\Gamma_{h}$ also increases \citep{Kubota18}. These relations have been clearly seen before \citep{Kubota18, Mitchell23}, but here the host galaxy subtraction also makes the warm Compton parameters more reliable. We see that there is a systematic increase in $\Gamma_{w}$, correlated with a decrease in $kT_{e, w}$, which implies that the warm Comptonising disc-like region is also systematically changing, from being consistent with all the luminosity being emitted in the photosphere ($\Gamma_{w} \sim 2.5$, a passive disc: \citealt{Petrucci18}), to being much steeper as expected for increasing dissipation in the warm disc itself.

Overall, our model fits show a clear and systematic evolution in the SED, which corresponds to a change in the accretion structure itself, transitioning from a disc(-like) dominated state at high $\mdot$ to one dominated by the X-ray plasma at low $\mdot$ (much like the low-hard state in BHBs.

\section{Discussion and Conclusions}
\label{sec:conc}

All the SEDs considered here are consistent with the standard outer disc component appearing below a roughly constant temperature of $\sim 4500 - 6500$\,K, which is also the temperature region where Hydrogen goes from being mostly ionised (above $T_{hi}$) to mostly neutral, perhaps providing insight into the origin of the warm Comptonising disc-like structure.

Disc temperatures around $1-5 \times 10^{4}$\,K predict local SEDs peaking in the UV/EUV. Stars at these temperatures (O stars, Wolf-Rayet, and luminous blue variables) have significant winds powered by UV line-driving (e.g \citealt{Puls08, ElMellah17}). This can translate to AGN discs, powering radiatively driven outflows, through e.g UV line driving \citep{Proga00, Proga04}. Strong winds from an accretion disc naturally lead to significant mass-loss, which can produce a distinct change in the disc properties themselves \citep{Laor14}. Our observed $T_{hi}$ is an order of magnitude below these characteristic UV wind temperatures, making \rev{the} warm disc region unlikely to be connected to this process.

Instead, there is a dramatic change in the disc opacity at 
the Hydrogen ionisation threshold at around $6000\,K$ \citep{Cannizzo92a}, leading to a local instability \citep{Cannizzo92b}. In stellar mass BHB discs, which cross this temperature at large radii where there is not much power from accretion, the disc opacity changes so dramatically that this local instability triggers a global limit cycle, \rev{within} the disc, so it transitions between outbursts and longer periods of quiescence \citep{Lightman74, King98, Lasota01, Dubus01}. The lower temperature discs in AGN mean that SMBHs cross this temperature range at much smaller radii (scaled in terms of the gravitational radius) where there is significantly more power emitted. Thus this instability might be expected to have an even more dramatic impact on the disc structure. However this is not generally seen in the simulations of this so far \citep{Burderi98, Siemiginowska97, Grzedzielski17},  but AGN discs are also in the regime where radiation pressure dominates, and we speculate that the non-linear outcome of these two instabilities acting together \rev{holds} the key to understanding the origin of the warm disc region in AGN.

\rev{In addition it is quite clear from Fig. \ref{fig:SED_fig} that while the optical luminosity changes by roughly 3 orders of magnitude, the hot X-ray component only appears to change by a factor of a few. While our model explains this as the flow becoming increasingly dominated by the X-ray corona as $\mdot$ decreases, we do not have a good physical reason for \emph{why} the flow seemingly smoothly transitions into an X-ray corona. Though, we note that the behaviour of the X-ray power is roughly consistent with the toy model of \citet{Lusso17}, which is based off \citet{Svensson94} (and expanded upon in \citealt{Merloni02} and \citealt{Arcodia19}) and states that the X-ray corona only becomes relevant once gas pressure begins to dominate over radiation pressure, setting a mass-accretion rate dependent size-scale of the X-ray corona. Alternatively, it appears that the X-ray always emits at $\sim 1-2$\,\% of the Eddington luminosity, which is also roughly the maximal luminosity for an ADAF flow \citep{Narayan95, Yuan14}. We speculate that perhaps this is showing that AGN always have a maximal ADAF present; contrary to BHBs.}

Regardless of the origin of the warm disc-like region \rev{and its transition into the X-ray corona}, it is fundamentally clear that it depends strongly on $\mdot$. The clear smooth transition in the SEDs from a disc dominated state to one dominated by the X-ray, is most easily explained by a change in the accretion structure itself, as shown previously for some rare changing-look (changing-state) AGN \citep{Noda18}. This change cannot be due to obscuration, as the X-rays are clearly unobscured (by selection \rev{- see also the discussion in Appendix \ref{app:Nh_test}}). The inner accretion flow in AGN is relatively compact, and so we would expect any obscuring medium present in the UV to intercept the X-rays too. Of course, there could be some low column-density dusty/gas on large scales that would contribute to the reddening of the optical/UV continuum, but would not be seen in our X-ray data (e.g \citealt{Czerny95}). However, we find it unlikely that this could cause sufficient reddening to give the extremely red continua seen in the lowest three luminosity bins, while remaining X-ray unobscured. In addition the transition from a strong soft X-ray excess in the highest three luminosity bins, to a single non-thermal X-ray component in the lowest three bins, strongly suggests a collapse of the structure responsible for the soft excess as $\mdot$ reduces rather than increased obscuration.

Along with the collapse of the disc, and thus the EUV continuum, we expect to see a reduction/loss of the broad emission lines, due to a reduction/loss of the ionising photons. This is entirely analogous with the changing-look phenomenon \citep{Noda18, Ruan19}, except now it is a generic feature of the AGN population rather than limited to a few rare objects. In some sense this has been previously hinted at, with studies showing a sharp drop in the number of broad line AGN below $0.02\,L_{\rm{Edd}}$ \citep{Trump11, Mitchell23}, which is both where the changing-look transition is known to occur and where our sample shows a loss of the EUV continuum. This can easily be tested by matching our HSC-eROSITA sample to one of the spectroscopic surveys (e.g SDSS \citealt{York00} or DESI \citealt{DESI16}). This is beyond the scope of our current paper, where we limit ourselves to the SED, but will be addressed in future work.

The more obvious limitation within our work is the data-gap between the optical/UV and the X-ray. Our modelling \emph{predicts} substantial power emitted in the EUV at high $\mdot$ and the loss of this power at low $\mdot$. However, this is predictive, and entirely based on what the model requires to bridge the gap between the optical/UV and X-ray, while maintaining energy conservation. It is entirely plausible that alternative accretion flow models could predict entirely different behaviour in the EUV gap, which would fit our data just as well. In this sense additional data in the near-UV (NUV) and far-UV (FUV) could work to strengthen (or falsify) our conclusions. This will be the topic of a future paper, using archival GALEX data (Kang et al. (in prep)).

Nonetheless, there is a clear drop in the blue optical continuum from the warm disc at $\mdot \lesssim 0.02$. This is clearly a generic feature in the AGN population, and not limited to local low-luminosity AGN (LINERs) \citep{Ho99} and a few rare individual changing-state AGN \citep{Noda18}. There is a real, and quite abrupt change in the accretion structure, driven by the changing Eddington ratio, which leads to the complete loss of the ionising EUV emission, and hence the characteristic broad emission lines. This has a significant impact on how we identify AGN through cosmic time. In an era where the James Webb Space Telescope is breaking new ground with the discoveries of quasars at cosmic dawn, any inference for the overall population requires we understand the nature of the accretion flow itself. This study reveals a systematic change in the accretion structure; the collapse of the accretion disc; between high and low luminosity AGN, impacting what we infer about the general AGN population, structure and growth through cosmic time.

\section*{Acknowledgements}

\rev{We would like to thank the referee Dr. Emanuele Nardini for useful comments which improved this manuscript. In particular for encouraging us to include the appendix with the additional mass-bins, strengthening the result.} SH acknowledges support from the Japan Society for the Promotion of Science (JSPS) through the short-term fellowship PE23722 and from the Science and Technologies Facilities Council (STFC) through the studentship grant ST/V506643/1. CD acknowledges support from STFC through grant ST/T000244/1 and Kavli IPMU, University of Tokyo. Kavli IPMU was established by World Premier International Research Center Initiative (WPI), MEXT, Japan. JS is supported by JSPS KAKENHI (23K22533) and the World Premier International Research Center Initiative (WPI), MEXT, Japan. This work was supported by JSPS Core-to-Core Program (grant number: JPJSCCA20210003).

\rev{This work is based on data from eROSITA, the soft X-ray instrument aboard SRG, a joint Russian-German science mission supported by the Russian Space Agency (Roskosmos), in the interests of the Russian Academy of Sciences represented by its Space Research Institute (IKI), and the Deutsches Zentrum für Luft- und Raumfahrt (DLR). The SRG spacecraft was built by Lavochkin Association (NPOL) and its subcontractors, and is operated by NPOL with support from the Max Planck Institute for Extraterrestrial Physics (MPE). The development and construction of the eROSITA X-ray instrument was led by MPE, with contributions from the Dr. Karl Remeis Observatory Bamberg \& ECAP (FAU Erlangen-Nuernberg), the University of Hamburg Observatory, the Leibniz Institute for Astrophysics Potsdam (AIP), and the Institute for Astronomy and Astrophysics of the University of Tübingen, with the support of DLR and the Max Planck Society. The Argelander Institute for Astronomy of the University of Bonn and the Ludwig Maximilians Universität Munich also participated in the science preparation for eROSITA.}

This work made use of the following {\sc python} modules: {\sc astropy} \citep{Astropy13, Astropy18, Astropy22} and {\sc numpy} \citep{Harris20}. Additionally, the plots were made using a combination of {\sc matplotlib} \citep{Hunter07} and {\sc seaborn} \citep{Waskom21}.

%%%%%%%%%%%%%%%%%%%%%%%%%%%%%%%%%%%%%%%%%%%%%%%%%%
\section*{Data Availability}

The optical HSC-fluxes used in this paper are from \citet{Li24}, and the corresponding catalogue can be 
found at: \href{https://member.ipmu.jp/john.silverman/HSC/}{https://member.ipmu.jp/john.silverman/HSC/}.

The X-ray spectra are part of the eROSITA eFEDS early data-release AGN catalogue, as described in \citet{Liu22}. This is available through: \href{https://erosita.mpe.mpg.de/edr/eROSITAObservations/Catalogues/}{https://erosita.mpe.mpg.de/edr/eROSITAObservations/Catalogues/}.



\bibliographystyle{mnras}
\bibliography{Refs} % if your bibtex file is called example.bib

%\begin{thebibliography}{99}
%\end{thebibliography}

%%%%%%%%%%%%%%%%%%%%%%%%%%%%%%%%%%%%%%%%%%%%%%%%%%

%%%%%%%%%%%%%%%%% APPENDICES %%%%%%%%%%%%%%%%%%%%%

\appendix

\section{Importance of the HSC host-AGN decomposition}
\label{app:hsc_decomp}

\begin{figure*}
    \centering
    \includegraphics[width=0.8\textwidth]{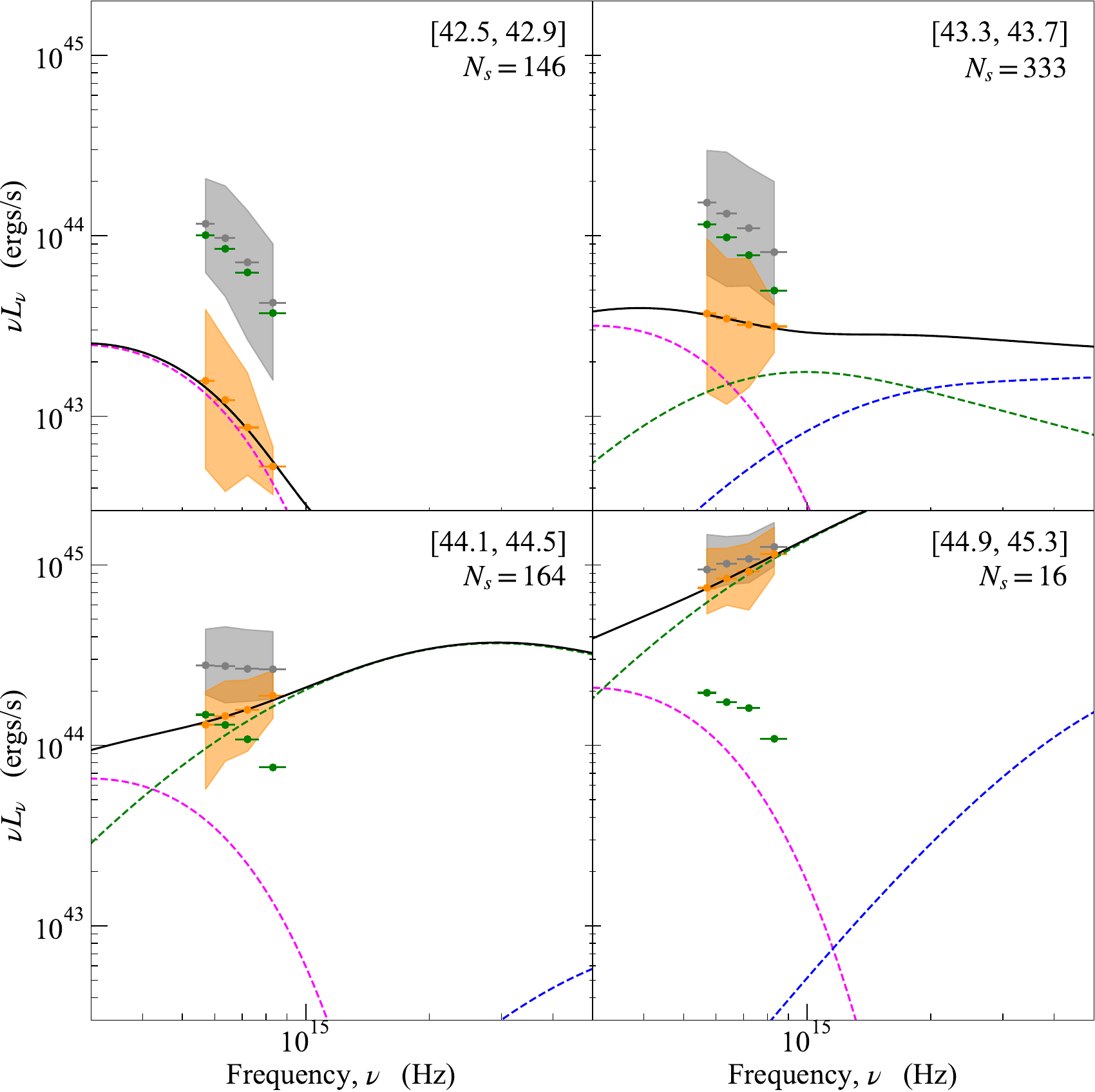}
    \caption{Zoom-in on the optical/UV region of the SEDs from Fig. \ref{fig:SED_fig}, for select luminosity bins (as indicated in the top right corners). Here we show the stacked data for the non-decomposed HSC fluxes (i.e host + AGN) in grey, with the decomposed stacks (AGN only) in orange, highlighting the necessity of the HSC decomposition. \rev{The green points show the galaxy only stacks, with the error region left off for clarity, to highlight that the evolution in the SEDs is driven by the AGN rather than the host.} In all, apart from the highest (likely super-Eddington), luminosity bins the inferred power and optical SED shape would be incorrect with no decomposition.}
    \label{fig:selectSEDs_incHost}
\end{figure*}

The results of our study rely on confidently extracting the AGN emission from the host galaxy emission, especially for the low luminosity bins. This is clear in Fig. \ref{fig:selectSEDs_incHost}, where we show the optical part of the stacked SEDs for both the de-composed AGN fluxes and the non-decomposed total fluxes for four of the luminosity bins \rev{($\nu L_{3500} \in [42.5, 42.9], \, [43.3, 43.7], \, [44.1, 44.5], \, \mathrm{and} \, [44.9, 45.3]$ ergs/s)}. In all panels, except the highest luminosity bin, we clearly see that the non-decomposed SED gives the wrong optical luminosity; from an error of a factor few to an order of magnitude. Hence an attempt at this study with no host galaxy decomposition would give the wrong estimate of the system energetics, over predicting the contribution from a disc (especially at low luminosity), and hence arriving at the wrong conclusion on the balance between the X-ray and optical emitting parts of the accretion flow.

As well as systematically overestimating the normalisation, we see in Fig. \ref{fig:selectSEDs_incHost} that the non-decomposed SEDs can give the wrong spectral shape (or slope). This is especially clear in the bin \rev{$\nu L_{3500} \in [43.3, 43.7]$\,ergs/s}, where the non-decomposed SED is clearly decreasing in power with frequency (energy), while the isolated AGN SED is more or less flat. 

The AGN and host galaxy fluxes are decomposed using a 2D optical image decomposition method, as described in \cite{Li24}. For details on how the method works we refer the reader to \cite{Li21a, Ding21}.

\rev{
\section{Critically testing for obscuration in the X-ray stacks}
\label{app:Nh_test}
}

\begin{figure}
    \centering
    \includegraphics[width=\columnwidth]{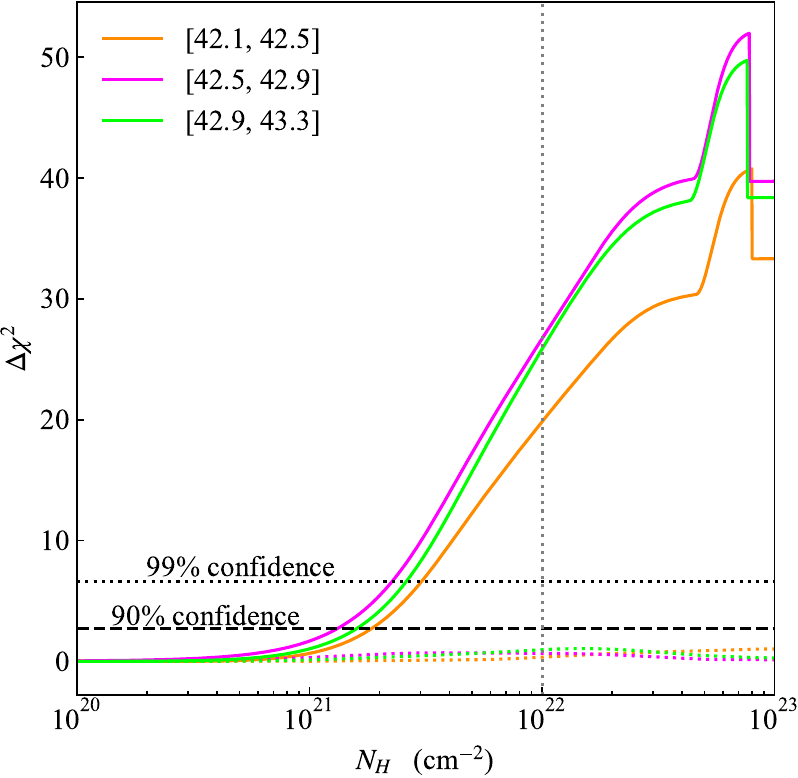}
    \caption{\rev{
    $\Delta \chi^{2}$ curves for an absorbed power-law model, using the X-ray stacks for the three lowest luminosity bins in Fig.
    \ref{fig:SED_fig}. The solid coloured lines show the case where the absorber covers all sources within the stacks (i.e {\sc phabs*pow}), while the dotted coloured lines show the case where only half the sources are obscured (i.e {\sc pcf*pow}). The horizontal dashed line and dotted line show the 90\,\% and 99\,\% confidence levels respectively, while the vertical dotted line highlights a column-density of $N_{H} = 10^{22}$\,cm$^{-2}$. It is clear that we completely rule out the X-ray stacks being obscured on average, however in the case where only half the sources are obscured we can make no upper constraints on $N_{H}$.
    }}
    \label{fig:Nh_step}
\end{figure}

\rev{
We have based our study on the assumption that all the sources within our stacks are in fact unobscured, as highlighted in our selection criterion of $N_{H, \rm{X-ray}} < 10^{22}$\,cm$^{-2}$. However, at this threshold $N_{H}$ and spectral slope will be degenerate in some of the eROSITA sources \cite{Liu22}, leading to uninformed $N_{H}$ measurements. Hence, there is a chance that some of the sources contributing to our stacks are in fact obscured. In the stacks presented in this paper (see Fig. \ref{fig:SED_fig}), this possibility mostly concerns the lowest three luminosity bins, which appear to be characterised by a single power-law. The higher bins cannot be heavily obscured on average, as they display a soft X-ray excess, which would be wiped out if the average column-density in these bins were above $10^{22}$\,cm$^{-2}$. In this appendix we therefore test the three lowest luminosity bins, to ensure that these results are convincingly due to a change in accretion state rather than spurious obscuration contaminating our sample.
}

\rev{
We start by taking the eROSITA stacks from the three lowest luminosity bins from Fig. \ref{fig:SED_fig} (i.e $\nu L_{3500} \in [42.1, 42.5], \, [42.5, 42.9], \, \rm{and} \, [42.9, 43.3]$). These initially appear as being characterised by a clear single power-law component. We apply a simple absorbed power-law model to these data, using the {\sc xspec} model {\sc phabs*pow}, to test maximally allowed average column-density in the sample. These results are shown in Fig. \ref{fig:Nh_step}, where we show $\Delta \chi^{2}$ curves calculated by stepping the fit through fixed $N_{H}$ values (solid coloured lines). It is immediately clear that on average the stacks are not obscured, ruling out $N_{H} \gtrsim 3 \times 10^{21}$\,cm$^{-2}$ at $99\,\%$ confidence.
}

\begin{figure}
    \centering
    \includegraphics[width=\columnwidth]{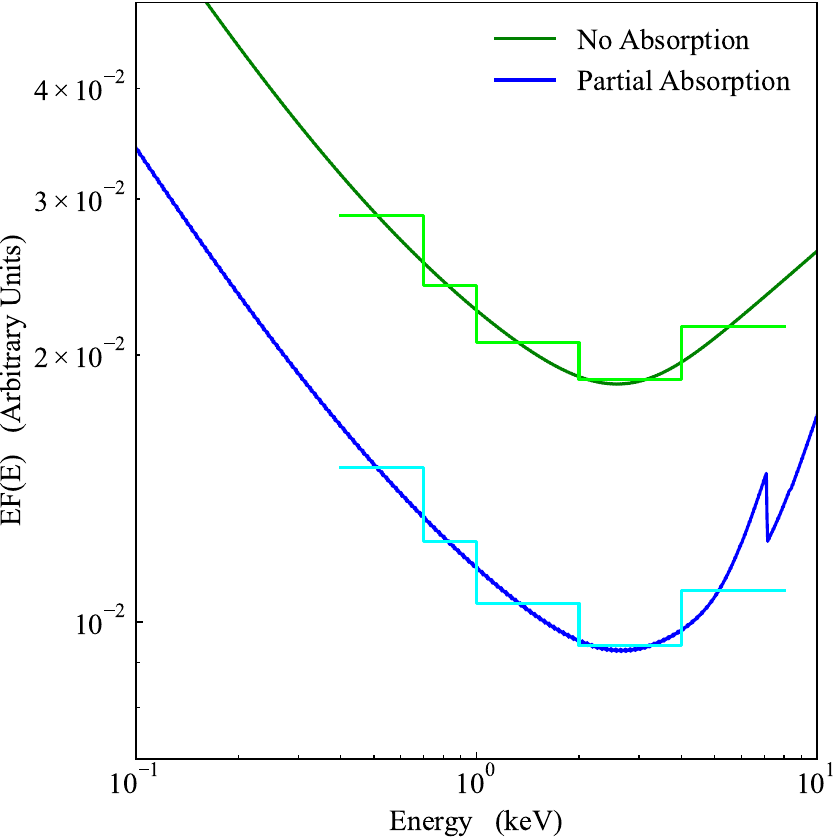}
    \caption{\rev{
    The X-ray portion of the model SED from luminosity bin $\nu L_{3500} \in [44.1, 44.5]$ from Fig. \ref{fig:SED_fig}. The solid green line shows the unabsorbed model, while the solid blue line shows the same model after including a 0.5 covering fraction absorber at $N_{H} = 10^{24}$\,cm$^{-2}$ in order to asses the impact of half the objects in a stack being heavily obscured. The stepped lines show the models on the energy grid of the eROSITA stacks.
    }}
    \label{fig:Nh_modComp}
\end{figure}

\rev{
However, the above is assuming all sources have some intrinsic obscuration. Hence, we now also test the case where only some of the sources are obscured, contaminating our stacks. Defining the {\sc xspec} model {\sc pcf*pow}, we set a covering fraction of 0.5, effectively assuming $50\,\%$ of the sources within the stacks are obscured. The results for this model are shown as the dotted coloured lines in Fig. \ref{fig:Nh_step}. In this case our stacks are clearly unable to constrain the column-density, with the results being consistent with both completely unobscured and half of the sources fully obscured. 
}

\rev{
To test whether a set of heavily absorbed sources within the stacks could feasibly turn a complex spectrum into what appears as a single power-law, we perform a test using the SED model from luminosity bin $\nu L_{3500} \in [44.1, 44.5]$ from Fig. \ref{fig:SED_fig}. Our choice falls on this bin as it give a relatively typical AGN SED, with a clear contribution from the soft X-ray excess, complicating the shape beyond a single power-law. Taking this SED model, we now apply the partial covering fraction absorption model {\sc pcf}, setting the covering fraction to 0.5 and column-density $N_{H} = 10^{24}$\,cm$^{-2}$ (i.e assuming half the sources within the stacks are heavily obscured). These results are shown in Fig. \ref{fig:Nh_modComp} where we can clearly see that, although the overall normalisation is altered, the spectral shape would not turn to a single power-law if half the objects within the stack are obscured. Certainly, the soft X-ray excess would still be observable within the stacks, leading us to conclude that the single power-law shape in our low-luminosity stacks must be intrinsic to the AGN themselves; not an artefact from spurious obscuration.
}

\rev{
\section{Alternate Mass Bins}
\label{app:mbin}
}

\begin{figure*}
    \includegraphics[width=\textwidth]{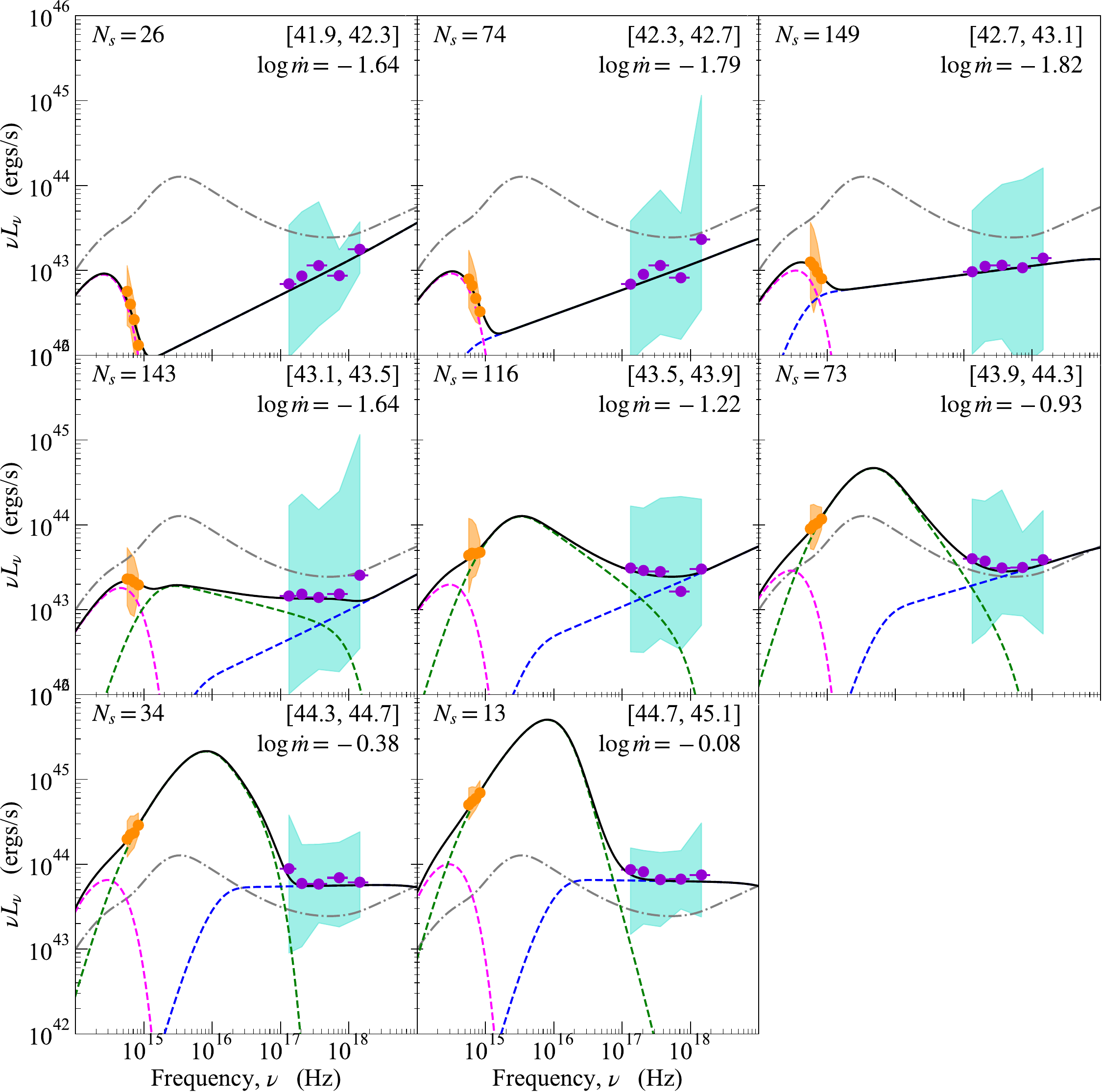}
    \caption{\rev{
    Same as Fig. \ref{fig:SED_fig}, but for the mass bin $\log \mmsun \in [7.5, 8.0]$
    }}
    \label{fig:SEDs_lowMbin}
\end{figure*}

\begin{figure*}
    \includegraphics[width=\textwidth]{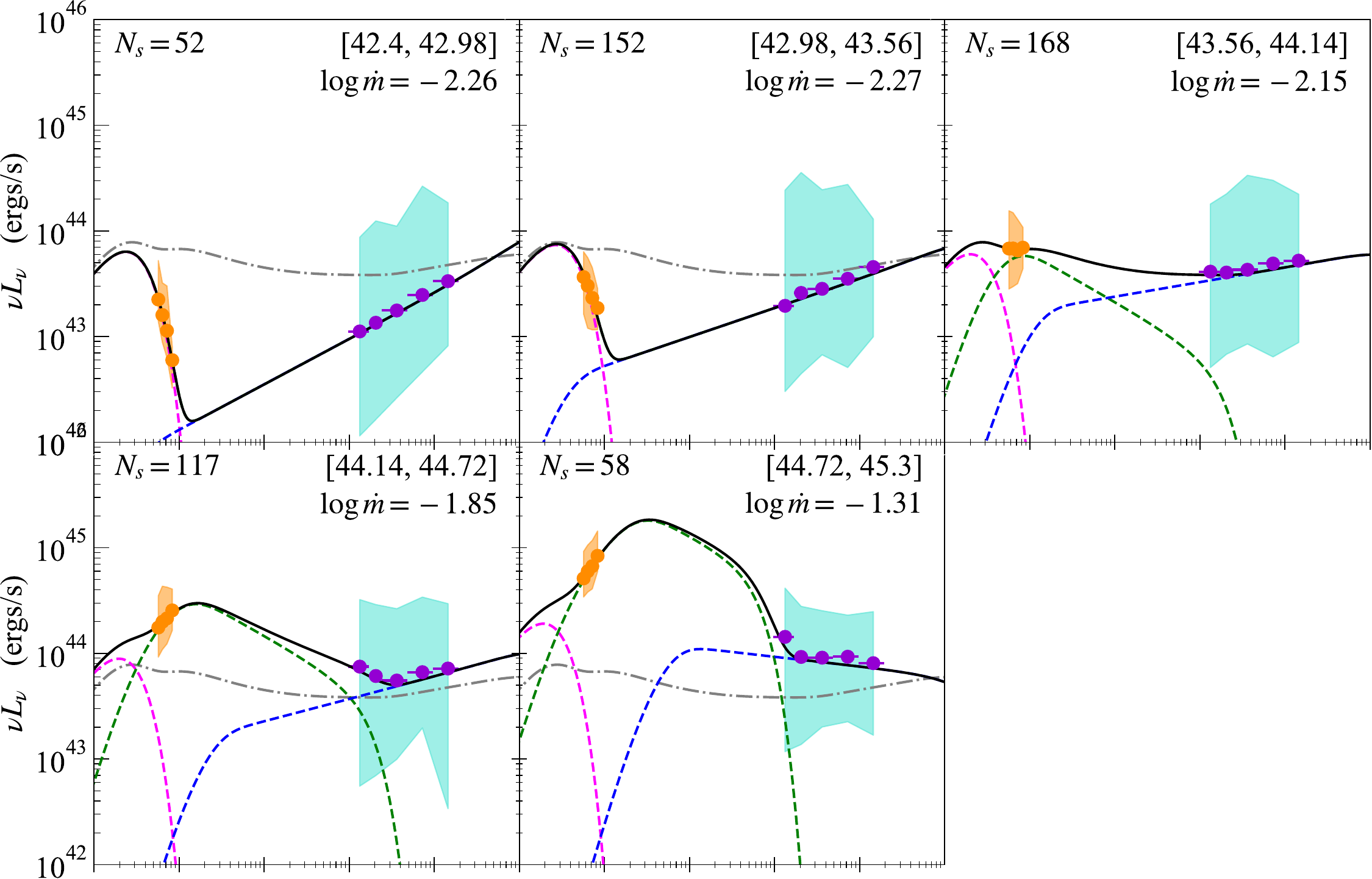}
    \caption{\rev{
    Same as Fig. \ref{fig:SED_fig}, but for the mass bin $\log \mmsun \in [8.5, 9.0]$
    }}
    \label{fig:SEDs_upMbin}
\end{figure*}

\rev{
Here we present the stacked SEDs for the mass bin below ($\log \mmsun \in [7.5, 8.0]$ - Fig. \ref{fig:SEDs_lowMbin} and above ($\log \mmsun \in [8.5, 9.0]$ - Fig. \ref{fig:SEDs_upMbin}) our fiducial mass bin, confirming that the same change in accretion state is seen beyond the single bin.
}

\rev{
We note that the luminosity binning differs between the mass bins, as we optimise this to give identical bin widths while also avoiding an uneven luminosity distribution in the edge bins. As with the fiducial mass bin, the stacks are too coarse to give well defined SED parameters. Nonetheless, the same trend is clear, with a receding disc as $\mdot$ reduces, eventually collapsing into a flow dominated by the X-ray plasma at low $\mdot$. The SED fit showing this are given in table \ref{tab:fitpars_altBins}.
}

\newpage
{\renewcommand{\arraystretch}{1.6} %Adjust table spacing
\begin{table*}
    \centering
    \begin{tabular}{c | c c c c c c c}
        $\log \nu L_{3500}$\,(ergs/s) & $\log \mdot$ & $kT_{e, w}$\,(keV) & $\Gamma_{h}$ & $\Gamma_{w}$ & $r_{\rm{hot}}$ & $T_{hi}$\,(K) & [$r_{w}(T_{hi})$] \\
        
        \hline 
        \hline 
        \multicolumn{8}{c}{$\log \mmsun \in [7.5, 8.0]$} \\
        \hline

        [41.9, 42.3] & -1.64 & $^\dagger$ & 1.60 & $^\dagger$ & $520^{\dagger \dagger}$ & 4913 & [520] \\
        
        [42.3, 42.7] & -1.79 & $^\dagger$ & 1.71 & $^\dagger$ & $351^{\dagger \dagger}$ & 6022 & [351] \\

        [42.7, 43.1] & -1.82 & $^\dagger$ & 1.90 & $^\dagger$ & $292^{\dagger \dagger}$ & 6719 & [292] \\

        [43.1, 43.5] & -1.64 & 0.97 & 1.62 & 2.20 & 55.3 & 8996 & [227] \\

        [43.5, 43.9] & -1.22 & 0.82 & 1.66 & 2.57 & 37.3 & 5913 & [557] \\

        [43.9, 44.3] & -0.93 & 0.57 & 1.77 & 3.08 & 18.8 & $5913^{*}$ & [701] \\

        [44.3, 44.7] & -0.38 & 0.07 & 1.99 & 2.90 & 11.6 & $5913^{*}$ & [1085] \\

        [44.7, 45.1] & -0.08 & 1 & 2.01 & 4.87 & 9.18 & $5913^{*}$ & [1398] \\ 

        \hline
        \hline
        \multicolumn{8}{c}{$\log \mmsun \in [8.5, 9.0]$} \\
        \hline

        [42.4, 42.98] & -2.26 & $^\dagger$ & 1.57 & $^\dagger$ & $157^{\dagger \dagger}$ & 4578 & [157] \\

        [42.98, 43.56] & -2.27 & $^\dagger$ & 1.73 & $^\dagger$ & $131^{\dagger \dagger}$ & 5209 & [131] \\

        [43.56, 44.14] & -2.15 & 0.19 & 1.86 & 2.45 & 74.4 & 3920 & [212] \\

        [44.14, 44.72] & -1.85 & 0.18 & 1.79 & 2.45 & 31.1 & $3920^{*}$ & 273 \\

        [44.72, 45.3] & -1.31 & 0.06 & 2.09 & 2.40 & 13.6 & $3920^{*}$ & [422] \\

    \end{tabular}
    \caption{\rev{
    Fit values for each of the luminosity bins (left column) for the stacked SEDs in the mass bin $\log_{10} \mmsun \in [7.5, 8.0]$ (top panel) and $\log_{10} \mmsun \in [8.5, 9.0]$ (bottom panel). The $r_{w}$ values in the square brackets (rightmost column) show the radius where the disc temperature reaches $T_{hi}$. \\
    $\dagger$ The warm Compton region is not required in this bin. \\
    $\dagger \dagger$ No warm Compton region is required so $r_{w}=r_h$, and $T_{hi}=T_{bb}(r_h)$. \\
    $*$ In these bins the standard outer disc is subdominant and $T_{hi}$ cannot be well constrained. Hence we freeze it to the last bin where it could make an impact (i.e $[43.9, 44.3]$ for $\log_{10} \mmsun \in [7.5, 8.0]$ and $[43.56, 44.14]$ for $\log_{10} \mmsun \in [8.5, 9.0]$ respectively.)
    }} 
    \label{tab:fitpars_altBins}
\end{table*}
}

%%%%%%%%%%%%%%%%%%%%%%%%%%%%%%%%%%%%%%%%%%%%%%%%%%

% Don't change these lines
\bsp	% typesetting comment
\label{lastpage}
\end{document}